\documentclass[nofootinbib,twocolumn,aps,prd,amsmath,amssymb,superscriptaddress]{revtex4}
\usepackage{graphics,bm}
\usepackage{epsfig}
\usepackage{graphicx}

\newcommand{\beq}{\begin{equation}}
\newcommand{\eeq}{\end{equation}}
\newcommand{\bqa}{\begin{eqnarray}}
\newcommand{\eqa}{\end{eqnarray}}

\newcommand{\unity}{1\hspace{-1.3mm}1}

\def\slashchar#1{\setbox0=\hbox{$#1$}     		
   \dimen0=\wd0                                 	
   \setbox1=\hbox{/} \dimen1=\wd1               	
   \ifdim\dimen0>\dimen1                        	
      \rlap{\hbox to \dimen0{\hfil/\hfil}}      	
      #1                                        	
   \else                                        	
      \rlap{\hbox to \dimen1{\hfil$#1$\hfil}}   	
      /                                         	
   \fi}      

\begin{document}

\parindent=20pt
\pagestyle{plain}
\font\tenrm=cmr10

\title{The phase diagram of neutral quark matter \\ with pseudoscalar
  condensates in the color-flavor locked phase}

\author{Harmen J. Warringa} \email{harmen@tf.phys.ntnu.no}
\affiliation{Department of Physics, Vrije Universiteit, De Boelelaan
  1081, 1081 HV Amsterdam, The Netherlands}
 \affiliation{Department of
  Physics, Norwegian University of Science and Technology, N-7491
  Trondheim, Norway}

\date{\today}

\begin{abstract}
  We calculate the phase diagram of color and electrically neutral
  quark matter in weak equilibrium using the NJL model at high baryon
  densities. We extend previous analyses by taking into account the
  possibility of formation of pseudoscalar meson condensates in the
  color-flavor locked (CFL) phase. We find that the kaon condensed
  phase (CFL-K$^0$) is preferred over the CFL phase at low
  temperatures. It is found that the CFL-K$^0$ phase contains no
  gapless modes. The results we have obtained should be of relevance
  for the modeling of compact stars.
\end{abstract}

\maketitle

\section{Introduction}
At sufficiently high baryon densities, massless three-flavor QCD is in
a color-flavor locked (CFL) phase \cite{Alford1999}. In the CFL phase
the original $\mathrm{SU}(3)_L \times \mathrm{SU}(3)_R \times
\mathrm{SU(3)_c} \times \mathrm{U}(1)_B$ symmetry of QCD is broken to
$\mathrm{SU}(3)_{V+c}$ due to the formation of scalar diquark
condensates.  As a consequence all eight gluons become massive by the
Anderson-Higgs mechanism. Furthermore the breakdown of chiral symmetry
leads to the emergence of eight massless pseudoscalar excitations,
similar to what happens in QCD at low baryon densities.  Casalbuoni
and Gatto have constructed a chiral effective theory for the CFL phase
\cite{Casalbuoni1999} that describes these massless excitations. In
such a chiral effective theory the pseudoscalar excitations $\pi^a$
are contained in a flavor octet chiral field $\Sigma = \exp(i \pi^a
\lambda^a /f_\pi)$.  Here $f_{\pi}$ is the pion decay constant and the
Gell-Mann matrices $\lambda_a$ are normalized as $\mathrm{tr} \,
\lambda_a \lambda_b = 2 \delta_{ab}$, where $a = 1 \ldots 8$.  The
field $\Sigma$ transforms under $\mathrm{SU}(3)_L \times
\mathrm{SU}(3)_R$ as $\Sigma \rightarrow U_L \Sigma U^\dagger_R$.

In the CFL phase with massless quarks the vacuum expectation value of
the $\Sigma$ field is equal to $1$. However, finite quark masses and
chemical potentials give rise to a potential for $\Sigma$
\cite{Son1999, Schafer2000, Bedaque2002}. Depending on the values of
the chemical potentials and the current quark masses $m_u$, $m_d$ and
$m_s$ the new ground state can have $\Sigma \neq 1$, which
arises via an axial flavor transformation on the CFL ground state.  In
such a case one of the pseudoscalar meson fields obtains a vacuum
expectation value \cite{Schafer2000}. Depending on the fields involved
one speaks of pion or kaon condensation giving rise to the
CFL-$\pi^\pm$, CFL-K$^0$ and CFL-K$^\pm$ phases \cite{Bedaque2002,
  Kaplan2002}.

Using the chiral effective theory it was found that pseudoscalar
condensation can occur if the chemical potential of the mesonic
excitation exceeds the mass of the corresponding excitation. For the
case $m_u = m_d$ the masses of the pionic and kaonic excitations (not
to be confused with the masses $m_\pi$ and $m_\mathrm{K}$ of the pion
and kaon mesons) in the chiral effective theory for the
CFL phase are respectively found to be \cite{Son1999} $M_\pi^2 = 2 a
m_u m_s$ and $M_{\mathrm{K}}^2 = a(m_u+m_s) m_u$. At asymptotically
high densities $a = 3 \Delta^2 / (\pi^2 f_\pi^2)$ \cite{Son1999},
where $\Delta$ is the CFL gap and the pion decay constant is given by
$f_\pi^2 = (21 - 8 \log 2) \mu^2 / (36 \pi^2)$.  Here $\mu$ denotes
one third of the baryon chemical potential.  The chemical potentials
of the pseudoscalar mesons are \cite{Kaplan2002} $\mu_{\pi^+} =
\mu_Q$,
\begin{multline}
\mu_{\mathrm{K}^+} = \mu_Q + \frac{m_s^2 - m_u^2}{2\mu},\;\;\;\;\;
\mu_{\mathrm{K}^0} = \frac{m_s^2 - m_d^2}{2\mu},
\label{eq:effchempots}
\end{multline}
where $\mu_Q$ is the chemical potential for electric charge. The
chemical potentials of the antiparticles are the opposite of those
above. The $\pi^0$ does not condense since its chemical potential
vanishes \cite{Kaplan2002}. The $\eta$ and the $\eta'$ excitation can
however obtain a vacuum expectation value \cite{Kryjevski2005}, but
will not be taken into account in this article.

The parameters of the chiral effective theory ana\-lysis have to be
matched to QCD, which only can be done accurately at extremely high
baryon densities where perturbative QCD is applicable. However, it is
interesting to know how matter behaves at non-asymptotic baryon
densities, such as inside a compact star. Therefore it would be
worthwhile to have an alternative analysis of pseudoscalar
condensation in the color superconducting phases which could be
reliable at intermediate baryon densities. The model we use for that
purpose here is the Nambu--Jona-Lasinio (NJL) model (see
\cite{Buballa2004a} for a review). The NJL model has been applied to
study the QCD phase diagram as a function of baryon chemical potential
and temperature under compact star constraints \cite{Ruster2005,
  Blaschke2005, Abuki2005} (see e.g.\ \cite{Weber2005} for a review of
compact stars).  In this article we will extend the NJL model analysis
of the phase diagram of Refs.\ \cite{Ruster2005, Blaschke2005,
  Abuki2005} by including the possibility of pseudoscalar condensation
in the color-superconducting phases. In this way we are able to obtain
a more detailed phase diagram for quark matter under compact star
constraints, that is electrically and color neutral matter in weak
equilibrium.  Earlier investigations of pseudoscalar condensation in
the CFL phase using the NJL model have been performed in Refs.\
\cite{Buballa2004b,Forbes2004}.  The phase diagram of quark matter
under compact star constraints was however not calculated in Refs.\
\cite{Buballa2004b,Forbes2004}. Reasonable agreement between the
chiral effective theory and the NJL model approach to pseudoscalar
condensation in the CFL phase was found in
\cite{Buballa2004b,Forbes2004}. This we will use as a justification to
apply some of the outcomes of the chiral effective theory analysis.

This article is organized as follows. In Sec.\ II we discuss shortly
the NJL model and the parameter choices. The details of the
calculation are given in Sec.\ III. The reader only interested in the
main results may go directly to Sec.\ IV in which the phase diagram of
quark matter under compact star constraints is presented. We summarize
and conclude in Sec.\ V.

\section{The NJL model}
In the NJL model, one treats the interaction between the quarks as a
point-like quark color current-current interaction. We will use the
following Lagrangian density
\begin{equation}
  \mathcal{L} = \bar \psi \left(i \gamma^\mu \partial_\mu - \hat m + \hat \mu
  \gamma_0 \right)\psi + \mathcal{L}_{\bar q q} + \mathcal{L}_{qq} +
  \mathcal{L}_\mathrm{H}, \label{eq:lagrnjl}
\end{equation}
where the quark-antiquark term, $\mathcal{L}_{\bar q q}$, the diquark
interaction term, $\mathcal{L}_{q q}$, and the 't Hooft interaction
term $\mathcal{L}_\mathrm{H}$ are defined below.  We have suppressed
the color (c), flavor (f), and Dirac (d) indices of the fermion fields $\psi$ for
notational simplicity.  The mass matrix $\hat m$ is diagonal and
contains the current quark masses. The matrix $\hat \mu$ is also
diagonal in flavor space and contains the quark chemical potentials.
In order to allow for electric and color neutralization \cite{Alford2002b, 
Colorneutral, Buballa2005b, Abuki2005} we write
\begin{equation}
  \hat \mu = \mu \unity_c \otimes \unity_f +  
  \mu_Q \unity_c \otimes\, Q +
  \mu_3 t_3 \otimes \unity_f +
  \mu_8 t_8 \otimes \unity_f,
\end{equation}
where $Q = \mathrm{diag}(2/3, -1/3, -1/3)$. When we later on in Sec.\
III we introduce a electron (e) and muon ($\mu$) gas, this choice will
automatically enforce the weak equilibrium conditions, $\mu_{d,s}
= \mu_u + \mu_{e,\mu}$. The color matrices corresponding to the color
chemical potentials $\mu_3$ and $\mu_8$ are $t_3 = \mathrm{diag}(1,
-1, 0)$ and $t_8 = \mathrm{diag}(1, 1, -2)/ \sqrt{3}$.  We use the
metric $g^{\mu \nu} = \mathrm{diag}(+\,-\,-\,-)$ and the standard
Dirac representation for the $\gamma$-matrices.  The quark-antiquark
interaction part of the Lagrangian density is
\begin{equation}
 \mathcal{L}_{\bar q q} 
  = G \left[ \left(\bar \psi \lambda_a \psi \right)^2 + 
 \left(\bar \psi \lambda_a i \gamma_5 \psi \right)^2 \right].
\end{equation}
Here the matrices $\lambda_a$ denote the $9$ generators of $\mathrm{U}(3)$ and
act in flavor space. They  
are normalized as $\mathrm{Tr}\, \lambda_a
\lambda_b = 2 \delta_{a b}$. To remind the
reader, the antisymmetric flavor matrices $\lambda_2$, $\lambda_5$ and
$\lambda_7$ couple up to down, up to strange and down to strange
quarks, respectively.
The diquark interaction term of the Lagrangian density is given by
\begin{eqnarray}
 \mathcal{L}_{qq} &=&   
  \frac{3}{4} G \left(\bar \psi t_{A} \lambda_{B} C i \gamma_5 \bar \psi^T \right)
\left(\psi^T  t_{A} \lambda_{B} C i \gamma_5 \psi \right)
  \nonumber \\
 && + \frac{3}{4} G \left(\bar \psi t_{A} \lambda_{B} C \bar \psi^T \right)
\left(\psi^T  t_{A} \lambda_{B} C \psi \right),
\end{eqnarray}
where $A, B \in \{2, 5, 7\}$ since only the interaction in the color
and flavor antisymmetric triplet channel is attractive.  The matrices
$t_a$ are the generators of $\mathrm{U}(3)$ and act in color space.
Their normalization is $\mathrm{tr}\, t_a t_b = 2 \delta_{a b}$.  The
charge conjugate of a field $\psi$ is denoted by $\psi_c = C \bar
\psi^T$ where $C = i \gamma_2 \gamma_0$.  The coupling strength $3 G
/4$ of the diquark interaction is fixed by the Fierz transformation
(see for example Ref.\ \cite{Buballa2004a}).  However, some authors discuss the NJL
  model with a different diquark coupling constant (in Ref.\
  \cite{Ruster2005} it was found that changing this coupling strength
  to $G$ can have a non-negligible effect on the phase diagram).
  Finally, the 't Hooft interaction which models an additional
  breakdown of the axial $\mathrm{U}(1)$ symmetry due to instantons is
  given by \cite{tHooft1976}
\begin{equation}
 \mathcal{L}_\mathrm{H} = -K \left [
  \mathrm{det}_f \bar \psi ( 1- \gamma_5) \psi
 +   \mathrm{det}_f \bar \psi ( 1 + \gamma_5) \psi
\right].
\end{equation}

The results that will be presented below are obtained with the
following choice of parameters \cite{Rehberg1995, Ruster2005}
\begin{eqnarray}
 m_{u} = m_{d} = 5.5\;\mathrm{MeV},\;\; m_{s} =
140.7\;\mathrm{MeV}
 \nonumber \\
 G = 1.835 / \Lambda^2, \;\;K = 12.36/\Lambda^5,\;\;\Lambda = 602.3\;\mathrm{MeV}.
\end{eqnarray}
This choice of parameters gives the following parameters the realistic
values \cite{Rehberg1995} $m_\pi = 135.0\;\mathrm{MeV}$, $m_{\mathrm{K}} =
497.7\;\mathrm{MeV}$, $m_\eta = 514.8\;\mathrm{MeV}$, $m_{\eta'} =
957.8\;\mathrm{MeV}$, and $f_\pi = 92.4\;\mathrm{MeV}$. Moreover, it
allows us to compare our results to the phase diagram obtained in
\cite{Ruster2005}, where only the scalar diquark condensates were
taken into account.

\section{Calculation of the phase diagram}
To obtain the phase diagram of the NJL model, we introduce 
the chiral condensates
\begin{equation}
  \sigma_u = \langle \bar u u \rangle, \;\;\;\; 
   \sigma_d = \langle \bar d d \rangle, \;\;\;\;
 \sigma_s  = \langle \bar s s \rangle,
\end{equation}
and the scalar and pseudoscalar diquark condensates
\begin{equation}
  \Delta^s_{AB} = \frac{3}{2}G \langle \psi^T t_A \lambda_B C \gamma_5
\psi \rangle,\;\;\;
  \Delta^p_{AB} =\frac{3}{2}G \langle \psi^T t_A \lambda_B C 
\psi \rangle.
\end{equation}
We do not take into account the possibility of pion condensation or
charged kaon condensation outside the CFL phase (e.g.\ the appearance
of $\langle \bar u i \gamma_5 d\rangle$ and $\langle \bar u i \gamma_5
s\rangle$, condensates).  Using the results of \cite{pionkaon} this is
in principle possible if $\vert \mu_Q \vert > m_\pi$ or $\vert \mu_Q
\vert> m_\mathrm{K}$.  Phase diagrams with these condensates have been
investigated in Refs.\ \cite{Barducci2004, Warringa2005, Ebert2005,
  He2006}.  From Ref.\ \cite{Ruster2005} we conclude that electric
neutralization gives values of $\vert \mu_Q \vert$ which are smaller
than $m_\mathrm{K}$ making the occurrence of this type of kaon
condensate unlikely.  Electric neutralization allows for values $\vert
\mu_Q \vert$ which are larger than $m_\pi$, but from the analysis of
Ref.\ \cite{Warringa2005} it follows that for matter in weak
equilibrium the phase with the pion condensate is not competing
against color superconducting phases.  This makes it in our opinion
improbable that such a condensate arises in neutral quark matter at
high baryon densities.

In the CFL phase with massless quarks the nonzero condensates are
$\Delta^s_{22} = \Delta^s_{55} = \Delta^s_{77}$. In the more realistic
situation of unequal quark masses and unequal chemical potentials the
same condensates appear, but then they are no longer equal to each
other. As was explained in the introduction the state with
pseudoscalar condensation in the CFL phase can be obtained by an axial
flavor transformation on the CFL state. In such a case
$\psi \rightarrow \exp(i \gamma_5 \theta_a \lambda_a / 2) \psi$ and the scalar
diquark condensates are partially rotated into pseudoscalar diquark
condensates \cite{Buballa2004b, Forbes2004}. For a K$^0$ mode the
diquark condensates transform as \cite{Buballa2004b}
\begin{eqnarray}
 \Delta^{s}_{22} \rightarrow \cos(\theta/2) \Delta^s_{22}, & 
   \Delta^{p}_{25} \rightarrow i \sin(\theta/2) (\hat \theta_6 - i \hat \theta_7)  \Delta^s_{22}, 
\nonumber \\
 \Delta^{s}_{55} \rightarrow \cos(\theta/2) \Delta^s_{55}, & 
   \Delta^{p}_{52} \rightarrow i \sin(\theta/2) (\hat \theta_6 + i \hat \theta_7) \Delta^s_{55}, 
\nonumber
\\
 \Delta^{s}_{77} \rightarrow \Delta^s_{77},\hspace{1.25cm} & 
 \label{eq:transformedcondensates}
\end{eqnarray}
where $\theta = \sqrt{\theta_6^2 + \theta_7^2}$ and $\hat \theta_a =
\theta_a / \theta$.  In the same way
$\Delta^s_{22}$ and $\Delta^s_{77}$ are partially rotated into
$\Delta^p_{27}$ and $\Delta^p_{72}$ under a K$^\pm$ transformation, 
while $\pi^{\pm}$ transformations
rotate $\Delta^s_{55}$ and $\Delta^s_{77}$ into $\Delta^p_{57}$ and
$\Delta^p_{75}$ \cite{Buballa2004b}. According to the analysis in the
chiral effective theory and confirmed numerically by Buballa
\cite{Buballa2004b} in the NJL model, the appearance of 
a pseudoscalar diquark condensate lowers the effective potential by
\cite{Kaplan2002}
\begin{equation}
 f_\pi^2 \mu_i^2 ( 1 - \cos \theta)^2 / 2,
 \label{eq:effpotlowering}
\end{equation}
where $i = \mathrm{K}^0$, $\mathrm{K}^\pm$ or $\pi^\pm$ 
and  $\cos \theta = M_i^2 / \mu_i^2$.

So the CFL-K$^0$ phase is characterized by the CFL condensates plus
two non-vanishing pseudoscalar diquark condensates $\Delta_{25}^p$ and
$\Delta_{52}^p$. Similarly, CFL-K$^\pm$ and CFL-$\pi^\pm$ phases exist.
Next to the CFL phase, different color superconducting phases are
known, like the 2SC \cite{2sc} (two-flavor color superconducting)
phase ($\Delta^s_{22} \neq 0, \Delta^s_{55} = \Delta^s_{77} = 0$) and
the uSC phase \cite{Neumann2003} ($\Delta^s_{22} \neq 0, \Delta^s_{55}
\neq 0, \Delta^s_{77} = 0$). In the phase diagram we will present, we
also encounter a phase with a $\Delta^s_{22}$ condensate together with
a non-vanishing $\Delta^p_{52}$. This phase we will call p2SC. A
nonzero value of $\Delta^p_{52}$ can be obtained by an axial color
transformation on the 2SC state. For convenience we have summarized
the possible combinations in Table \ref{tb:condensates}.

\begin{table}
\begin{tabular}{c|ccccccccc}
phase & 
$\Delta^s_{22}$ & 
$\Delta^s_{55}$ &
$\Delta^s_{77}$ &
$\Delta^p_{25}$ &
$\Delta^p_{52}$ &
$\Delta^p_{27}$ &
$\Delta^p_{72}$ &
$\Delta^p_{57}$ &
$\Delta^p_{75}$ \\
\hline
\hline 
2SC & $\times$ & 0 & 0   &   0 & 0   &   0 & 0    &  0 & 0 \\
p2SC & $\times$ & 0 & 0 &   0 & $\times$  &   0 & 0    &  0 & 0 \\
uSC & $\times$ & $\times$ & 0   &   0 & 0   &   0 & 0    &  0 & 0 \\
CFL & $\times$ & $\times$ &  $\times$  &   0 & 0   &   0 & 0    &  0 & 0 \\
CFL-K$^0$ & $\times$ & $\times$ &  $\times$  &    $\times$ & $\times$   &   0 & 0    &  0 & 0 \\
\hline
CFL-K$^\pm$ & $\times$ & $\times$ &  $\times$  &    0 & 0  &   $\times$  & $\times$     &  0 & 0 \\
CFL-$\pi^\pm$ & $\times$ & $\times$ &  $\times$  &   0 & 0  &   0 & 0    &  $\times$  & $\times$  \\
dSC  & $\times$ & 0 &  $\times$   &    0  & 0   & 0 & 0     &  0 & 0 \\
sSC & 0 & $\times$ & $\times$   &    0  & 0  & 0 & 0     &  0 & 0 \\
2SCus  & 0 & $\times$ & 0   &    0  & 0  & 0 & 0     &  0 & 0 \\
2SCds & 0 & 0 & $\times$   &   0  & 0   & 0 & 0     &  0 & 0 \\
\end{tabular}
\caption{Definition of different phases. A $0$ or a $\times$ denotes respectively
  a vanishing or non-vanishing condensate. If at
  least one of the quasi-particles has a gapless mode one writes a 'g' in front
  of the name of the phase, e.g.\ g2SC, gCFL, etc. If
  $\mu_Q > 0$, one speaks of CFL-K$^+$ and CFL-$\pi^+$, if  $\mu_Q < 0$ of
  CFL-K$^-$ and CFL-$\pi^-$.
  The five phases above the horizontal line appear in the phase diagram of
  neutral quark matter, Fig.~\ref{fig:mubt}. The other phases are
  possibilities (among others) which we have taken into account, but did not arise
  in the phase diagram.}
\label{tb:condensates}
\end{table}

To obtain the phase diagram of the NJL model under compact star
constraints with the possibility of pseudoscalar condensation in the
color-superconducting phases we will proceed as follows. In order to
allow for electric neutralization, we introduce a free electron and
muon gas and calculate the finite temperature ($T$) mean field
effective potential which yields (see for example \cite{Ruster2005,
  Buballa2004b})
\begin{multline}
  \mathcal{V} = 2 G (\sigma_u^2 + \sigma_d^2 + \sigma_s^2)
+ \frac{\left \vert \Delta^s_{AB}\right \vert^2 
  + \left \vert \Delta^p_{AB}\right \vert^2 }{3 G}
\\ 
- 4 K \sigma_u \sigma_d \sigma_s 
-\frac{T}{2} \sum_{p_0 = \tilde \omega_n} \int \frac{\mathrm{d}^3 p}{
\left( 2\pi \right)^3} \log \mathrm{det} S^{-1} 
\\
- \frac{T}{\pi^2} \sum_{l=e, \mu} \sum_\pm \int_0^\infty \mathrm{d} p \, p^2
\log \left[ 1 + e^{-\beta( E_l \pm \mu_Q )} \right],
\label{eq:effpotential}
\end{multline}
where $\tilde \omega_n = (2 n + 1) \pi T$ and $E_l = \sqrt{p^2 +
  m_l^2}$. Weak equilibrium is automatically satisfied in this way
(for example one can read off that $\mu_s = \mu_u + \mu_e$).  Here we
have assumed that neutrinos leave the star immediately. An analysis
with neutrino trapping, as done for the phase diagram with scalar
diquark condensates in Ref.\ \cite{Ruster2006}, is left for future
work.  The electron and muon mass have the following values, $m_e =
0.511\;\mathrm{MeV}$ and $m_\mu = 105.66\;\mathrm{MeV}$. The $72
\times 72$ matrix $S^{-1}$ contains the inverse Nambu-Gorkov fermion
propagator in the mean field approximation
\begin{equation}
S^{-1} = 
\left(
\begin{array}{cc}
\unity_c \otimes \mathcal{D} 
+ \hat \mu \otimes \gamma_0 
&
t_A \otimes \lambda_B \otimes \Phi^1_{AB}
\\
t_A \otimes \lambda_B \otimes \Phi^2_{AB}
&
\unity_c \otimes \mathcal{D}
-
\hat \mu \otimes \gamma_0  
\end{array} \right).
\end{equation}
Here 
\begin{equation}
  \mathcal{D} = \unity_f \otimes (i \gamma_0 p_0 + \gamma_i p_i) 
 - \hat M \otimes \unity_d 
\;,
\end{equation}
and
$\hat M = \mathrm{diag}(
m_{0u} - 4 G \sigma_u + 2 K \sigma_d \sigma_s, \;
m_{0d} - 4 G \sigma_d + 2 K \sigma_s \sigma_u,\;
m_{0s} - 4 G \sigma_s + 2 K \sigma_u \sigma_d)$.
The matrices $\Phi^1_{AB}$ and $\Phi^2_{AB}$ are equal to
\begin{eqnarray}
 \Phi^1_{AB} &=& \Delta^{s}_{AB}  \gamma_5 + \Delta^{p}_{AB} \unity_d,
\\
 \Phi^2_{AB} &=& -\Delta^{s*}_{AB} \gamma_5 + \Delta^{p*}_{AB} \unity_d.
\end{eqnarray}
The matrix $\unity$ is the identity matrix in color ($c$), flavor ($f$),
or Dirac ($d$) space.

To calculate the effective potential in an efficient way, one can
multiply the matrix $S^{-1}$ with $\mathrm{diag}(\unity_c \otimes
\unity_f \otimes \gamma_0, \unity_c \otimes \unity_f \otimes \gamma_0
)$ which leaves the determinant invariant.  In this way, one obtains a
new matrix $R$ with $i p_0$'s on the diagonal. We can write $R = i p_0
\unity + A$, where $A$ is a Hermitian matrix independent of $p_0$.  By
determining the eigenvalues of the matrix $A$, one can reconstruct the
determinant of $S^{-1}$ for all values of $p_0$ which is 
$\prod_{i=1}^{72}(\lambda_i + ip_0)$. After summing over Matsubara
frequencies, one finds
\begin{equation}
 T \!\!\!\sum_{p_0 = \tilde \omega_n} \!\!\! \log
  \mathrm{det} S^{-1} = \sum_{i=1}^{72} \left[\frac{\vert \lambda_i \vert}{2} + T
  \log\left(1 + e^{-\vert \lambda_i \vert / T} \right) \right].
\end{equation}
All that remains in order to determine the effective potential is to
perform the integration over three-momentum $p$ up to the ultraviolet
momentum cutoff $\Lambda$ numerically.

In order to enforce electric and color charge neutrality the following
equations have to be satisfied 
\begin{equation}
 n_Q = - \frac{\partial \mathcal{V}}{\partial \mu_Q} = 0,\;\;\;\;
 n_{3,8} = - \frac{\partial \mathcal{V}}{\partial \mu_{3,8}} = 0
 \label{eq:chargeneutr}
\end{equation}
where $n_Q$ is the electric charge density, and $n_3=n_r-n_g$, and
$n_8=(n_r+n_g-2n_b)/\sqrt{2}$. 

The values of the condensates and the phase diagram are determined by
minimizing the effective potential ${\cal V}$ with respect to the
set of possible condensates. Together with the electric and charge
neutrality constraints this implies that we have to solve the following
equation
\begin{equation}
 \frac{\partial \mathcal{V}}{\partial x_i} = 0,
 \label{eq:effpotminimization}
\end{equation}
where $x = \left\{\sigma_u, \sigma_d, \sigma_s, \Delta^s_{AB}
\Delta^p_{AB}, \mu_Q, \mu_3, \mu_8 \right\}$.

We will allow for the chiral condensates and the possible diquark
condensates denoted in Table \ref{tb:condensates}.  The analysis in
the effective theory shows that the different phases with pseudoscalar
condensation are separated by a first-order transition
\cite{Kaplan2002}. Therefore it is reasonable to minimize the
effective potential with respect to the different possible sets of
pseudoscalar diquark condensates separately. Thereafter one can
compare the minimal values of the effective potential in order to
determine the phase. This speeds up the numerical calculation.

Using the diagonal color transformations and the $\mathrm{U}(1)$
flavor transformations it is always possible to choose the scalar
diquark condensates to be real, the phases of the pseudoscalar
condensates however cannot be rotated away. We find that the effective
potential of the NJL model with {\it real} CFL condensates and two
pseudoscalar diquark condensates $\Delta^p_{25}$ and $\Delta^p_{52}$
is invariant under the transformation $\Delta^p_{25} \rightarrow e^{i
  \phi} \Delta^p_{25}$ together with $\Delta^p_{52} \rightarrow e^{-i
  \phi} \Delta^p_{52}$ (for the other possible sets of diquark
condensates, the same transformation rule of course holds). This shows
that one can always rotate one diquark condensate to be real. From
Eq.\ (\ref{eq:transformedcondensates}) it follows that in the minimum of
the effective potential the other diquark condensate is then real as
well.  We have also checked this numerically, by allowing for a
complex pseudoscalar diquark condensate. Hence we can take the three
scalar and the two pseudoscalar diquark condensates to be real. This
speeds up the calculation as well. 

To evaluate the derivatives of the effective potential (see also
\cite{Abuki2005}) we have used the following formula
\begin{multline}
T \frac{\partial}{\partial x_j}
 \sum_{p_0 = \tilde \omega_n}  \log
  \mathrm{det} S^{-1}(x) \\
= \frac{1}{2} \sum_{i=1}^{72} b_i 
\left(1 - \frac{2}{e^{\vert \lambda_i \vert / T} + 1} \right) 
 \mathrm{sgn}(\lambda) \;,
 \label{eq:effpotderiv2}
\end{multline}
where $b_i = (U^\dagger \partial A/\partial x_j U)_{ii}$. Here
$\lambda_i$ are again the eigenvalues of $A$, and $U$ is a unitary
matrix which contains in the $i$-th column the normalized eigenvector
of $A$ with eigenvalue $\lambda_i$.  To obtain the complete derivative
of the effective potential one still has to integrate
Eq.~(\ref{eq:effpotderiv2}) over momenta up to the cutoff $\Lambda$.
The advantage of using Eq.~(\ref{eq:effpotderiv2}) to evaluate the
derivatives is threefold: firstly it is more accurate than the finite
difference method, secondly it is also much faster since one had to
diagonalize the matrix $A$ anyway in order to obtain the value of the
effective potential, and finally additional derivatives can be
computed without much extra numerical work.

The speed of the calculation of the effective potential and its
derivatives depends heavily on how fast one can compute the
eigenvalues. There are several ways to speed up the calculation.
Firstly, the determinant of $A$ does not depend on the direction of
$\vec p$. Therefore, one can choose $\vec p$ to lie in the
$z$-direction.  Together with the choice of the non-vanishing
condensates mentioned above, this implies that $A$ becomes a real
symmetric matrix, which simplifies the calculation of the eigenvalues.
Secondly, one can interchange rows and columns of $A$ without changing
its determinant. By doing so, one can bring $A$ in a block-diagonal
form. One can then determine the eigenvalues of the blocks separately
which is significantly faster since the time needed to compute
eigenvalues numerically scales cubically with the dimension of the
matrix. In the most general case with two pseudoscalar diquark
condensates, one can always reduce the problem to one $20 \times 20$
matrix and three $8 \times 8$ matrices.

We determined the eigenvalues using LAPACK routines \cite{Lapack1999}.
The numerical integration over three-momentum $p$ up to the cutoff was
done using a Gauss-Legendre quadrature with $2 \times 32$ points at $T
> 15\;\mathrm{MeV}$. For lower temperatures the Bose-Einstein
distribution functions start to behave like step-functions, therefore
the Gauss-Legendre quadrature, which is based on polynomial
interpolation, becomes less accurate. For that reason we used Simpson
integration with 256 points at lower temperatures. We have checked the
accuracy of the integration procedures by doubling the number of
integration points by 2. The condensates were determined by solving
Eqs.  (\ref{eq:effpotminimization}) and (\ref{eq:chargeneutr}) using a
multidimensional rootfinding algorithm of the GSL package \cite{gsl}
with different initial conditions. The ground state was determined by
picking the solution that has the lowest value of the effective
potential. After the ground state was found we investigated the zeros
of $\vert \lambda_i \vert$ as a function of momentum. If $\vert
\lambda_i \vert$ has a zero, one of the quasi-particles has a gapless
mode. In that case one speaks of the gCFL, g2SC, or guSC phase
\cite{gapless}. The problem of gapless phases are its chromomagnetic
instabilities \cite{gaplessinstabilities}. Therefore such a phase
cannot be the phase with the lowest free energy.

\section{Results}
In this section we will discuss the phase diagram of quark matter
under compact star constraints, that is electrically and color neutral
matter in weak equilibrium. We will discuss two possible checks of our
results.  The first is to compare with the results of the chiral
effective theory. The other is to compare with the phase diagram of
Ref.\ \cite{Ruster2005} in which no pseudoscalar color-superconducting
phases were taken into account.

We first present the phase diagram in the $\mu_B$-$\mu_Q$ plane (so no
electric neutralization) at $T=0$.  This allows for comparison with
the chiral effective theory calculations of \cite{Kaplan2002}.  
Thereafter we will discuss the phase diagram of neutral quark matter.

\subsection{The $\mu_B$-$\mu_Q$ phase diagram at $T=0$}
To predict the shape of the $\mu_B$-$\mu_Q$ phase diagram one can use
Eqs.\ (\ref{eq:effchempots}) and (\ref{eq:effpotlowering}). The
resulting phase diagram is somewhat different from the analysis in the
chiral effective theory of Ref.\ \cite{Kaplan2002}, because here
electromagnetic corrections to the masses of the charged mesonic
excitations have not been taken into account.  Firstly, if $\mu_Q$ is
positive, $\mu_{\mathrm{K}^+} > \mu_{\mathrm{K}^0}$ and
$\mu_{\mathrm{K}^+} > \mu_{\pi^+}$. Since the energy gain of meson
condensation is proportional to $\mu_i^2 - 2M^2_i + M_i^4 / \mu_i^2$
and $\mu_i > M_i$, it follows that the CFL-K$^+$ phase is favored over
the CFL-K$^0$/$\pi^+$ phase for positive values of $\mu_Q$. For
negative values of $\mu_Q$ the CFL-K$^0$ phase is favored over the
CFL-K$^-$ phase, because then $\mu_{\mathrm{K}^-} <
\mu_{\mathrm{K}^0}$.  So there has to be a phase transition from the
CFL-K$^0$ phase to the CFL-K$^+$ phase at $\mu_Q = 0$. Secondly, if
$\mu_Q$ is negative enough, pion condensation will be favored over
neutral kaon condensation. Since $M_\mathrm{K}$, $M_\pi$ and
$\mu_{\mathrm{K}^0}$ are all proportional to $1 / \mu$ the transition
line should be proportional to $1 / \mu$ as well at asymptotically
high densities.

\begin{figure}[t]
\includegraphics{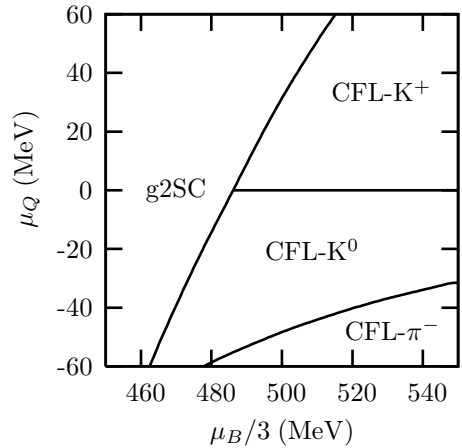}
\caption{The phase diagram of color neutral quark matter in weak
  equilibrium, as a function of baryon chemical potential and the
  electric charge chemical potential for $T=0$. The solid lines denote
  first-order phase transitions.\label{fig:mubmuq}}
\end{figure}

These conclusions are in agreement with the calculation of the phase
diagram presented in Fig.\ \ref{fig:mubmuq}. For $\mu_Q >0$ we find
the CFL-K$^+$ phase. There is a first-order phase transition at $\mu_Q
= 0$. At negative $\mu_Q$ we encounter first the CFL-K$^0$ phase and
then the CFL-$\pi^-$ phase. If the baryon chemical potential is
increased, the transition from the CFL-K$^0$ to the CFL-$\pi^-$ phase
occurs at smaller values of $\mu_Q$, in agreement with the
predictions. 

We want to stress that, in order to calculate this diagram, we have
used that chiral effective theory analysis \cite{Kaplan2002} shows
that the phase boundaries between the the phases with pseudoscalar
condensation are first-order. Therefore, in order to simplify the
calculation, we have assumed that this holds in the NJL model as well.

The phase diagram of color neutral matter in weak equilibrium as
presented in Fig.\ \ref{fig:mubmuq}, does not necessarily contain all
phases which are present in the phase diagram of matter that is also
electrically neutral. This is because the electric neutrality
constraint can force the electrically neutral system to be in a state
which is a meta-stable state of color neutral matter in weak
equilibrium.

\subsection{The phase diagram of neutral quark matter}

\begin{figure*}[t]
\includegraphics{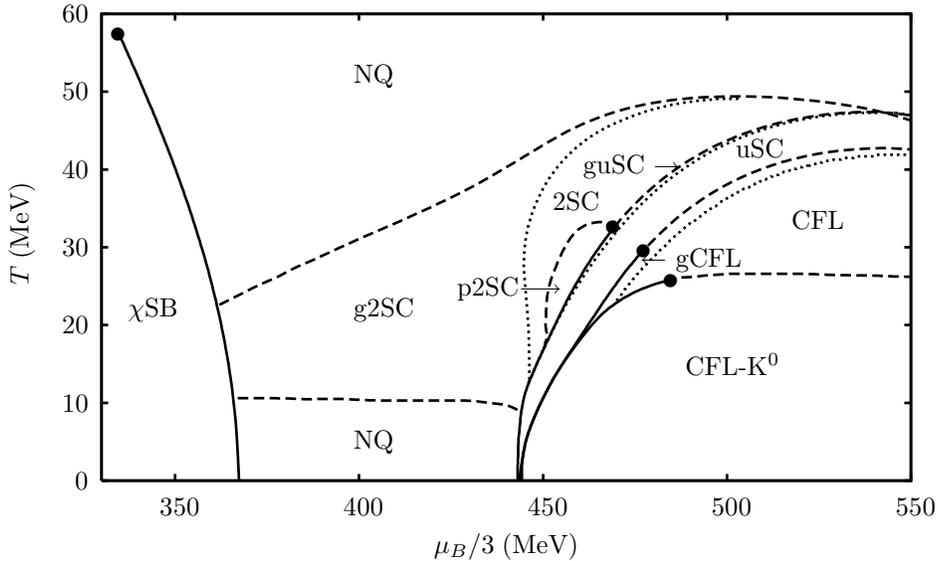}
\caption{The phase diagram of electrically and color neutral quark
  matter in weak equilibrium, as a function of baryon chemical
  potential and temperature. first-order phase transitions are denoted 
  by a solid line, second-order phase transitions by a dashed line.
  The dotted lines indicate the (dis)appearance of gapless modes in
  the quasi particle excitation spectrum. Critical endpoints are denoted
  by a black dot. The $\chi$SB and the NQ labels denote the chirally 
  broken and neutral chirally symmetric quark matter phases respectively. 
  The definition of the color-superconducting phases can be found in 
  Table \ref{tb:condensates}.
  \label{fig:mubt}}
\end{figure*}

In Fig.\ \ref{fig:mubt} we have displayed our main result, the phase
diagram of quark matter under compact star constraints, that is color
and electrically neutral matter in weak equilibrium. This phase
diagram was obtained by minimizing the effective potential on a grid
of 440 (in the $\mu_B$ direction) by 300 (in the $T$ direction)
points. In this way we can determine the phase boundaries up to an
accuracy of $\pm 0.25\;\mathrm{MeV}$ in the $\mu_B$ direction and $\pm
0.1\;\mathrm{MeV}$ in the $T$ direction. After we minimized the
effective potential on this grid we have checked the continuity of the
minimum value of the effective potential.

\begin{center}
\begin{figure*}
\hspace{0.3cm}$T = 0\;\mathrm{MeV}$ \hspace{3.5cm}
$T = 20\;\mathrm{MeV}$ \hspace{3.2cm}
$T = 40\;\mathrm{MeV}$ \\
\hspace{-2.0cm}\begin{tabular}{rrr}
\includegraphics{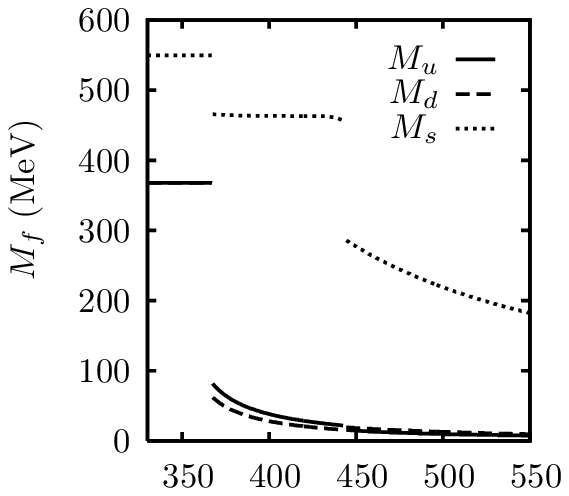} &
\includegraphics{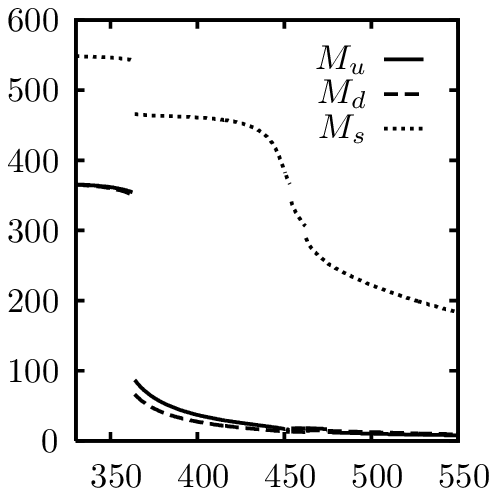} &
\includegraphics{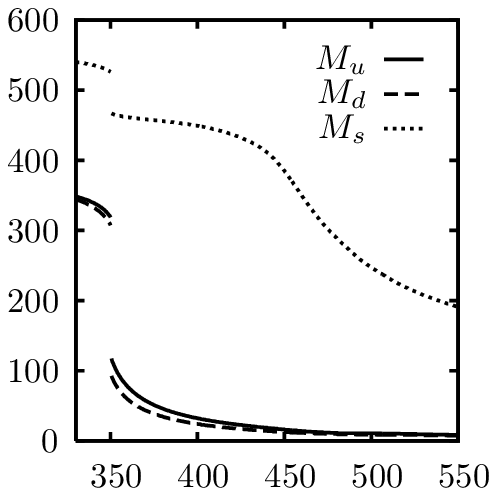} \\

\includegraphics{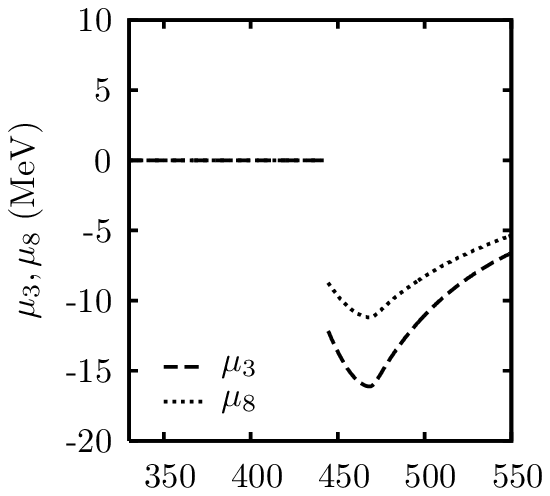} &
\includegraphics{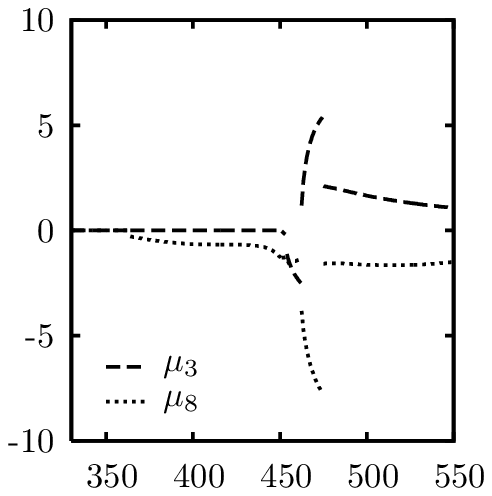} &
\includegraphics{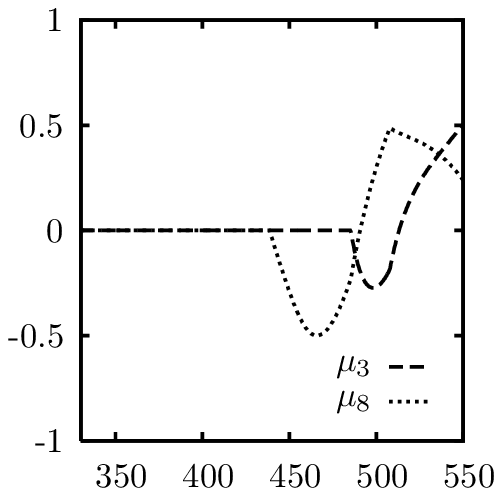} \\

\includegraphics{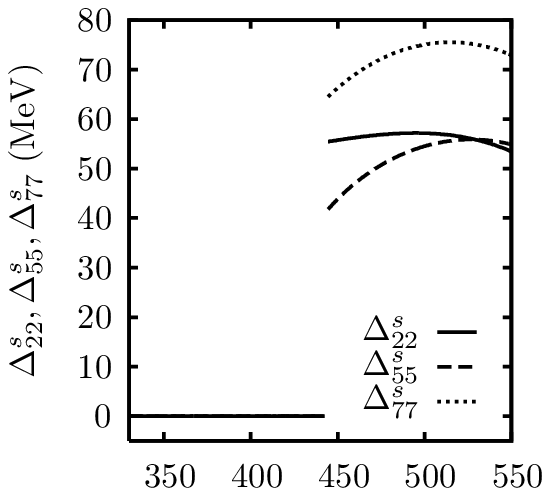} &
\includegraphics{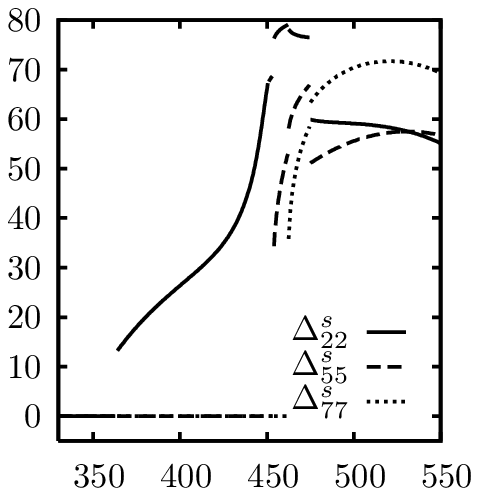} &
\includegraphics{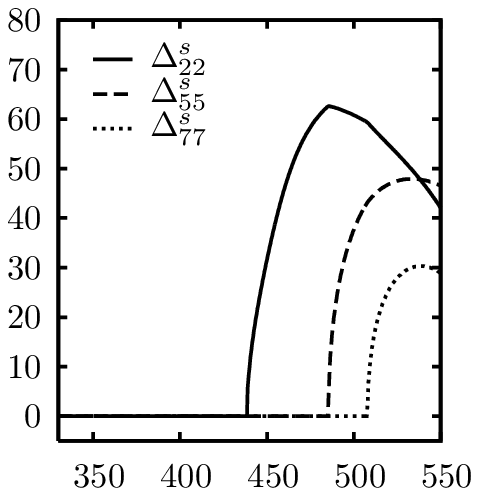} \\

\includegraphics{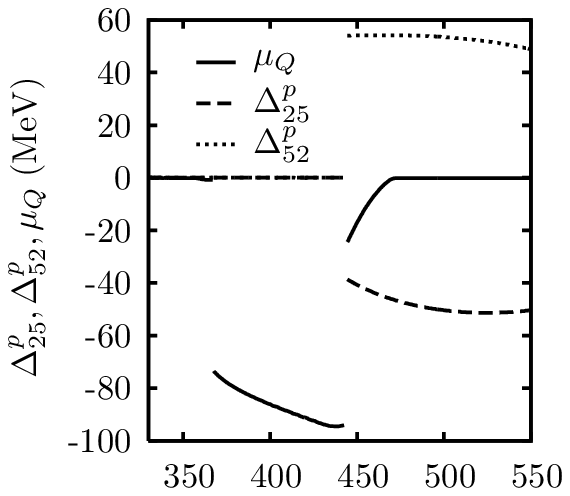} &
\includegraphics{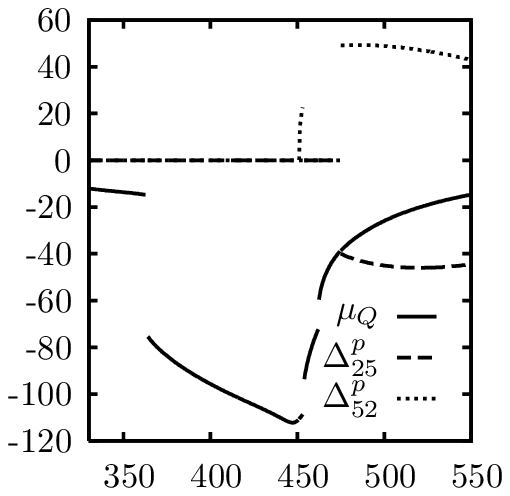} &
\includegraphics{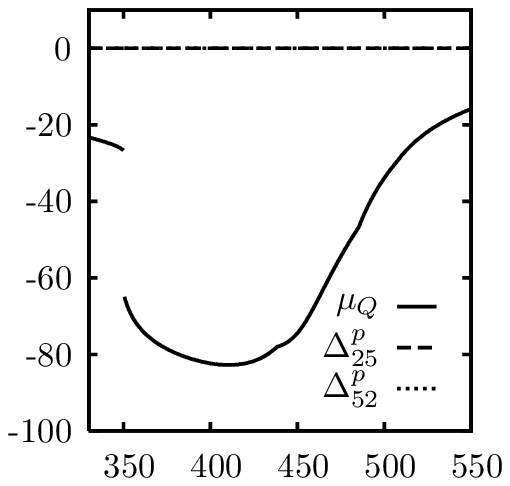} 
\end{tabular}\\
\hspace{0.8cm}$\mu_B / 3\;(\mathrm{MeV})$ \hspace{3.3cm}
$\mu_B / 3\;(\mathrm{MeV})$ \hspace{3.3cm}
$\mu_B / 3\;(\mathrm{MeV})$ 
\\
\caption{Constituent quark masses, chemical potentials and condensates
  for electrically neutral quark matter in weak equilibrium. The
  constituent quark masses, chemical potentials and condensates are
  displayed as a function of baryon chemical potential, for $T
  =0\;\mathrm{MeV}$ (left-hand side), $T=20\;\mathrm{MeV}$ (middle
  panels) and $T =40\;\mathrm{MeV}$ (right-hand side).\label{fig:condens}
 }
\end{figure*}
\end{center}

In Fig.\ \ref{fig:condens} we have displayed the values of the
condensates and the chemical potentials in the minimum of the
effective potential for $T=0\;\mathrm{MeV}$, $T=20\;\mathrm{MeV}$ and
$T=40\;\mathrm{MeV}$. Outside the CFL phase, excellent agreement is
found with \cite{Ruster2005} (their $\mu_3$ and $\mu_8$ differ by a
factor 2 from ours due to another normalization). Since we took the
pseudoscalar condensates to be real, a characteristic feature can be
read off from Eq.\ (\ref{eq:transformedcondensates}), namely that they
should have opposite sign. This we have indeed found and can be seen
in our results in Fig.\ref{fig:condens}.

The most important conclusion to be drawn from the phase diagram is
that (for our realistic parameter choices) electrically and color
neutral quark matter in weak equilibrium is at high baryon densities
and low temperatures in the CFL-K$^0$ phase. We find that this phase
contains no gapless modes, which implies that it does not contain
chromomagnetic instabilities. 

The CFL-K$^0$ is not always favored over the CFL phase. If at high
baryon densities the temperature is raised the CFL-K$^0$ phase goes
via a second-order phase transition to the CFL phase. At lower baryon
densities we find a critical point where the transition to the CFL
phase becomes first-order.

In electrically neutral quark matter, there should always be an excess
of electrons and muons to compensate for the lower density of strange
quarks compared to up and down quarks. Therefore $\mu_Q$ has to be
necessarily negative (this can be seen in Fig.\ \ref{fig:condens}
as well), which implies that only the CFL-K$^0$ and the CFL-$\pi^¯$
phase could in principle be found in electrically neutral quark matter.
We find the CFL-K$^0$ phase, but we do not encounter the CFL-$\pi^-$ phase 
in the phase diagram.

\begin{figure}[t]
\includegraphics{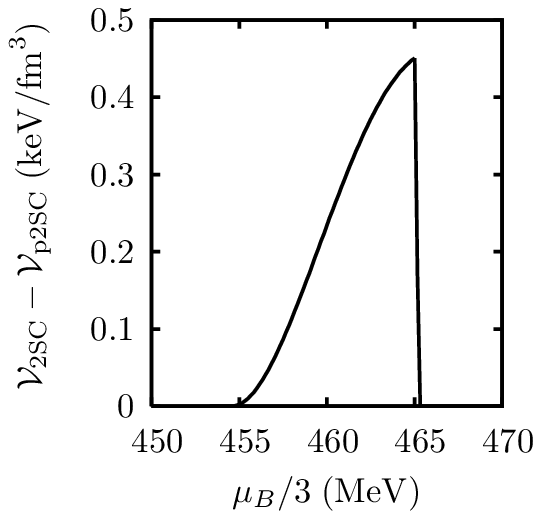}
\caption{The difference between the minimal values of the effective
  potential of the meta-stable 2SC phase and the p2SC phase for $T=30\;\mathrm{MeV}$.
  \label{fig:effpot}}
\end{figure}

Another striking feature of the phase diagram is the occurrence of the
p2SC phase, a two-flavor color super conductor with a non-vanishing
pseudoscalar diquark condensate, $\Delta_{52}^p$. We find that the
difference in free energy between the p2SC phase and the 2SC phase is
exceptionally small. While usually a difference in the minimal value
of the effective potential between two competing phases is in the
order of $0.1 - 1\;\mathrm{MeV} / \mathrm{fm}^3$, in this case it is
only of order $0.1\; \mathrm{keV} / \mathrm{fm}^3$, see Fig.\
\ref{fig:effpot}.  Therefore it could well be that other parameter
choices will let the p2SC phase disappear, but nevertheless it is very
interesting that its free energy is so close to the ordinary 2SC
phase.

\begin{figure}[t]
\includegraphics{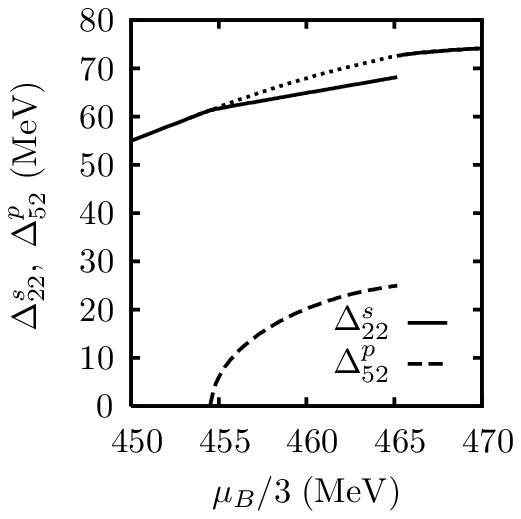}
\caption{The values of the condensates in the p2SC phase
for $T=30\;\mathrm{MeV}$. The
dotted line indicates the meta-stable 2SC solution.
\label{fig:p2sccond}}
\end{figure}

We have displayed the values of the condensates of the p2SC phase in
Fig.\ \ref{fig:p2sccond} again for $T=30\;\mathrm{MeV}$. From this
figure one can see that upon increasing the baryon chemical potential,
one enters the p2SC phase via a second-order phase transition. The
dotted line in the figure indicates the $\Delta^s_{22}$ condensate of
the meta-stable 2SC phase. One can see from Fig.\ \ref{fig:p2sccond}
that the $\Delta^p_{52}$ condensate has a relatively small value
compared to the $\Delta^s_{22}$ condensate.  We find that the square
root of the sum of the values of the $\Delta^s_{22}$ and the
$\Delta^p_{25}$ condensates is equal to value of the meta-stable
$\Delta^s_{22}$ condensate. This shows that the p2SC phase can be
obtained from the 2SC phase by an axial color transformation.

In our definition the p2SC phase contains a non-vanishing
$\Delta_{22}^s$ and a $\Delta_{52}^p$ condensate. Using the
off-diagonal color rotations, one can rotate a $\Delta_{52}^p$ in a
$\Delta_{72}^p$ condensate. We have checked that taking into account a
$\Delta_{22}^s$ and $\Delta_{72}^p$ condensate leads to the same phase
diagram. We have also investigated whether other possible combination
of $\Delta_{22}^p$, $\Delta_{52}^p$ and $\Delta_{72}^p$ lead to a
lower free energy, but we did not find such solutions with lower free
energy.

We have also sought other possible combinations of two pseudoscalar
diquark condensates and three scalar diquark condensates (of which one
or more could potentially vanish), but we did not find any of them.

Comparing our results to the phase diagram of \cite{Ruster2005} (where
the CFL-K$^0$ phase was not considered) we have good agreement outside
the CFL-K$^0$ and the p2SC phase. The phase boundaries of the 2SC
phase coincide. Also we find the boundaries of the g2SC and the guSC
gapless phases at the same place. At higher temperatures this also
holds for the CFL and the gCFL phase boundaries (at lower temperatures
we find the CFL-K$^0$ phase).  Inspection shows that at zero
temperature at $\mu = 443\;\mathrm{MeV}$ there is a first-order
transition from the neutral quark matter phase to the CFL-K$^0$ phase.
This transition point is in excellent agreement with the calculations
of \cite{Ruster2005} where at the same point a transition to the CFL
phase was found. A difference is that we find that there is a very
small window for the uSC phase, even for $T<5\;\mathrm{MeV}$.
Furthermore, our critical endpoints lie at somewhat higher
temperatures.

It must be said that other sensible choices of parameters can lead to
different phase diagrams (see e.g.\ \cite{Ruster2005} for a
comparison). We however believe that the CFL-K$^0$ phase should be
relatively stable against changing parameters. This does probably not
hold for the p2SC phase, because of the very small difference in free
energy between the p2SC and the 2SC phase. One must also be aware that
in reality also crystalline phases \cite{Alford2001b} and phases with
color-spin locking (see e.g.\ \cite{Aguilera2005}) are possible. Also
effects of non-local interactions could potentially modify the phase
structure \cite{GomezDumm2005}. At low baryon densities, nuclear
effects become important, which should also be taken into account (see
e.g.\ \cite{Lawley2006} for a unified description of nuclear and quark
matter in the NJL model).

Let us finally indicate some astrophysical consequences of the phase
diagram displayed in Fig.\ \ref{fig:mubt} compared to the phase
diagram obtained in \cite{Ruster2005} in which no CFL-K$^0$ phase was taken
into account. In \cite{Ruster2005} the CFL-K$^0$ phase is replaced by
a CFL and a gCFL phase. The difference in free energy between the
CFL-K$^0$ and the CFL phase is not large, so the new phase may not
have a huge influence on the equation of state and hence the
mass-radius relationships of a compact star.  However, the neutrino
cooling of a star is very sensible to the low-energy excitations
\cite{Schafer2004}. In phases without gapless modes, like the
CFL-K$^0$ phase we found, the neutrino emissivity is exponentially
suppressed compared to in phases with gapless modes \cite{Schafer2004}.
Hence replacing the gCFL by the CFL-K$^0$ phase will reduce the
cooling speed of a star with a quark core.

\section{Conclusions}
In this paper, we have calculated the $\mu_B$ vs.\ $T$ phase diagram
of electrically and color neutral quark matter in weak equilibrium, at
high baryon densities. It could be useful for understanding
compact stars with a quark matter core.

Our analysis has shown that neutral quark matter is in the CFL-K$^0$
phase at high baryon densities and low temperatures, rather thane the
CFL phase. Other CFL phases with pseudoscalar condensation were not
found.  This CFL-K$^0$ phase is different from the ordinary CFL phase,
since in the CFL-K$^0$ phase two pseudoscalar diquark condensates
appear.  The CFL-K$^0$ phase we obtained, contains no gapless modes.
This implies that there are no chromomagnetic instabilities in the
CFL-K$^0$ phase and that the neutrino emissivity is exponentially
suppressed in the CFL-K$^0$ phase.

The CFL-K$^0$ phase does not completely replace the CFL phase. At
higher temperatures we found a first and a second-order phase
transition to the CFL phase.

Next to the CFL-K$^0$ phase, we found at intermediate densities and
temperatures a 2SC phase which contains a pseudoscalar diquark
condensate, which we have called p2SC. This phase can be obtained
by an axial color rotation on the 2SC phase. The difference in the free
energy between the 2SC and the p2SC phase is very small. This implies
that corrections may let this phase disappear. In any case it would
be interesting to have a more detailed investigation of this phase.

We have obtained good agreement with earlier calculations of the phase
diagram were pseudoscalar condensation was not taken into account and
with results from the chiral effective theory. 

\section*{Acknowledgments}
I would like to thank D.\ Boer and J.O.\ Andersen for useful
discussions. Furthermore, I would like to thank D.~Boer for valuable
comments on the manuscript. The numerical calculations were performed
on the Dutch National compute cluster Lisa, which is maintained by
SARA computing and networking services.

\end{document}